# Quantitative tracking of grain structure evolution in a nanocrystalline metal during cyclic loading

Jason F. Panzarino, Jesus J. Ramos, Timothy J. Rupert[*]

Department of Mechanical and Aerospace Engineering, University of California, Irvine, CA 92697, USA

[*]E-mail: trupert@uci.edu

**Abstract**

Molecular dynamics simulations were used to quantify mechanically-induced structural evolution in nanocrystalline Al with an average grain size of 5 nm. A polycrystalline sample was cyclically strained at different temperatures, while a recently developed grain tracking algorithm was used to measure the relative contributions of novel deformation mechanisms such as grain rotation and grain sliding. Sample texture and grain size were also tracked during cycling, to show how nanocrystalline plasticity rearranges overall grain structure and alters the grain boundary network. While no obvious texture is developing during cycling, the processes responsible for plasticity act collectively to alter the interfacial network. Cyclic loading led to the formation of many twin boundaries throughout the sample as well as the occasional coalescence of neighboring grains, with higher temperatures causing more evolution. A temperature-dependent cyclic strengthening effect was observed, demonstrating that both the structure and properties of nanocrystalline metals can be dynamic during loading.

1. **Introduction**

As average grain size (*d*) is reduced to the nanometer range to create a nanocrystalline material, traditional intragranular dislocation mechanisms become suppressed [1, 2]. The higher stress levels which can be accessed and the increased grain boundary (GB) volume fraction in these materials exaggerates the importance of interfaces as plasticity mechanisms begin to depend heavily on the grain boundary network. As average grain size is reduced to the smallest possible nanocrystalline regime (*d* < 15-20 nm), plasticity becomes dominated by phenomena such as GB sliding [3-5], grain rotation [6-8], and GB migration [9-11]. Molecular dynamics (MD) studies have proven invaluable for exploring these novel deformation mechanisms. For example, Keblinski et al. [12] and Yamakov et al. [13] simulated grain-boundary (Coble) diffusion creep in nanocrystalline metals, providing evidence that the process is accommodated by Lifshitz type boundary sliding. The mechanical response of individual boundaries has also been explored extensively. For example, Tucker and McDowell [14] simulated bicrystals subjected to tension and shear and found that the local atomic structure of GBs can influence nanocrystalline deformation mechanisms. Kumar et al. [15] observed strain accommodation in notched NC aluminum specimens through GB shuffling, grain rotation, and Ashby-Verrall type grain switching, which reduced stress concentrations at the notch tip. The effect of free surfaces on the yield behavior of nanocrystalline and nanotwinned Cu nanowires was highlighted by Wu et al. [16], who showed that GB sliding can become more important in nanowire samples. The complexity of nanocrystalline plasticity can even include coupling of multiple mechanisms as shown by Upmanyu et al. [17], who observed a simultaneous activation of grain rotation and boundary migration.

A common feature of these new mechanisms is that they all represent collective motion of groups of atoms and should alter the internal grain structure of the material. For example, rotation of two neighboring grains would alter the misorientation between the grains and the character of the GB which separates them. Prior investigations of nanocrystalline plasticity have shown that strain accommodation mechanisms can occur cooperatively throughout the microstructure, leading to evolution of the interfacial network through large scale sliding and rotation processes. For example, Hasnaoui et al. [18] used molecular dynamics simulations to observe the emergence of shear planes in nanocrystalline Ni through the alignment of grain boundary planes via grain rotation, GB sliding, and GB migration, which then resulted in localized shear deformation as well as grain coalescence and growth. These observations were supported by the model of Hahn and Padmanabhan [19], which explains that substantial sliding on the mesoscale can only be accomplished by long range interconnection of interfaces. A recent MD study by Rupert [20] also showed that strain localization in nanocrystalline metals is highly dependent on the connectivity of the interfacial network. For the finest nanocrystalline grain sizes, the high strain path was formed through only grain rotation and GB sliding, with no significant evidence of dislocation motion. Once neighboring grains align to form a percolating GB network, catastrophic shear deformation occurs suddenly. In contrast to coarse grained metals whose plastic deformation is governed by intragranular interactions between dislocations, nanocrystalline metals have exhibited structural evolution such as mechanically-induced grain growth [21] during loading.

The importance of grain structure evolution should be exaggerated during low cycle fatigue loading, where multiple cycles of plastic strain are applied. It has been suggested that cyclic softening or strengthening processes during fatigue are highly dependent on the initial microstructure of the nanocrystalline metal [22]. Nanocrystalline Ni with minimal initial defect densities exhibited cyclic hardening during fatigue tests [23, 24]. Moser and colleagues were also able to identify a shift in plasticity from dislocation based to GB mediated during the initial fatigue cycles of nanocrystalline Ni [24]. Rupert and Schuh [25] observed a strengthening effect during MD simulations of cyclic loading in nanocrystalline Ni, which they ascribed to boundary relaxation and the formation of low energy boundaries. Grain growth has also been found to occur during mechanical cycling, as shown by tensile load-unload fatigue tests of nanocrystalline Cu by Witney et al. [26]. Schiotz also observed grain coarsening using molecular dynamics of Cu subjected to large tension-compression strain cycles [27]. While these studies show that nanocrystalline grain structures are dynamic, most characterization has been relatively qualitative and an opportunity exists for the quantification of grain structure evolution.

Because of the extremely small length scales associated with nanocrystalline materials, it can be difficult to observe and even more challenging to quantify structural evolution using experimental techniques. In-situ transmission electron microscopy (TEM) has been helpful in confirming the transition to GB mediated plasticity for these materials, and the direct observation of grain rotation [7, 28, 29] and GB migration [10] can be found in the literature. However, most of these results are qualitative in nature since it is often difficult to track the orientation of individual nanograins, especially for the finest grain sizes where GB mechanisms dominate. Kobler et al. [29] have recently overcome such limitations by straining Au films inside of the TEM while generating nanoscale crystal orientation maps at several stages of deformation. Such a study allowed these authors to quantify grain growth and orientation changes during the mechanical loading experiment. Unfortunately, the resolution of the technique used by Kobler et al. is limited and they restricted their data analysis to grain sizes above 15 nm. In addition, such TEM orientation maps give two dimensional maps of the microstructure. Since a grain's rotation and sliding with respect to its neighbors is likely a complicated three dimensional process, techniques which can track structural changes in all spatial dimensions would be extremely useful.

Fortunately, atomistic simulations provide materials researchers with the required spatial and temporal resolution needed to document quantitative structural evolution trends in nanostructured materials. Since the exact positions of all atoms are known, attributes such as crystallinity, bond structure, and orientation can be readily calculated [30] and visualized [31] throughout the atomistic simulation. These metrics can be tracked during the simulation and used to identify trends in structural evolution. In addition, post-processing tools have recently been developed which are capable of detecting and analyzing features such as dislocation networks [32] and grain structure [33] .

With previous work providing evidence that these collective GB processes are taking place, quantification of structural evolution during plasticity is the next logical step in furthering our understanding of nanocrystalline deformation physics. In this paper, we use MD to simulate the cyclic deformation of nanocrystalline aluminum, with an emphasis on understanding how grain structure evolves during loading. During cycling at three different temperatures, we quantify the amounts of grain rotation, grain growth, and grain sliding using a recently developed tool for identifying grains and tracking their properties. These mechanisms are found to be very temperature-dependent, with higher testing temperatures leading to more collective reorganization of the grain structure. Significant structural changes are observed, with the most notable being the formation of many twin boundaries in the material.

## 2. Computational Methods

Computational nanocrystalline Al samples were made using the Voronoi tessellation construction, a well-established method for simulating nanocrystalline specimens and studying their properties [34-37]. Euler angles were randomly generated for each grain nucleation site in order to obtain a random texture distribution throughout the sample and any overlapping atoms were removed once grains were grown. MD simulations were performed with the open-source Large-scale Atomic/Molecular Massively Parallel Simulator (LAMMPS) [38] code with an integration time step of 2 fs and an embedded atom method (EAM) potential for Al developed by Mishin et al. [39]. MD time steps must be lower than the shortest vibrational period of the atoms in the solid (typically ~1 ps) [40] and commonly used time steps vary from 1-5 fs. The validity of a 2 fs time step was verified by comparing with an identical mechanical cycling simulation at 300 K using a smaller time step of 1 fs. The cyclic stress-strain curves were virtually identical, but the longer time step allowed for faster simulation. The Al structure was first relaxed using a conjugate gradient minimization in LAMMPS with an energy tolerance of $10^{-6}$ eV and force tolerance of $10^{-6}$ eV/Å. This approach resulted in a fully dense nanocrystalline Al sample with $d = 5$ nm and starting grains free of stored dislocations. Nanocrystalline thin films have similarly been shown to be dislocation-free in their as-deposited condition [41]. The initial configuration contained 44 grains and periodic boundary conditions were imposed in all directions to simulate a repeating volume element of nanocrystalline Al. Previous work has shown that local GB structure can be in a non-equilibrium state, one which contains excess boundary dislocations or free volume [14]. This non-equilibrium structure can be "relaxed" during annealing or deformation [25, 42], and the boundary can reach a lower energy state. While non-equilibrium boundary structures are present in as-prepared nanocrystalline MD samples, we want to focus on more obvious changes to the grain structure here. Therefore, we anneal our samples at 600 K before testing in order to remove those non-equilibrium boundaries that may be a byproduct of our sample construction technique. The samples were equilibrated at 600 K for 100 ps to relax GB structure, and then cooled at a rate of 30 K/ps until the desired testing temperature was reached. Cyclic loading simulations were run at 600 K, 450 K, and 300 K, to investigate the effect of temperature on deformation physics and structural evolution.

Monotonic tension and tensile load-unload cyclic loading was applied along the Z-axis of the samples. A constant true strain rate of $5 \times 10^8$ s$^{-1}$ was used and an isothermal-isobaric (NPT) ensemble was employed to keep zero stress in the lateral sample directions. Monotonic loading was conducted by straining the nanocrystalline Al to 10% true strain. Low cycle fatigue was simulated by first loading to 5% strain to induce plasticity, and then unloading and reloading in 2% strain increments. The cycling began after the yield point of the sample, so that microstructural changes due to plastic strain accommodation could be observed. In other words, the sample was unloaded to 3% and then reloaded to 5% to complete one fatigue cycle. Up to 10 cycles were performed, with full loading to 10% true strain after 5 and 10 cycles to investigate the effects of cycling on subsequent mechanical behavior. Atomic positions were analyzed to measure important structural features after the initial 5% strain (cycle 0) as well as at the conclusion of each unloading-loading cycle (also 5% strain). This provided several data sets for each temperature from which evolution trends could be extracted.

The strain rate of $5 \times 10^8$ s$^{-1}$ was chosen to balance simulation time with sample size, with this value lying within the range of strain rates reported to be acceptable in the MD literature. Previous computational work [4, 43, 44] and experimental work [45-47] has reported that the mechanical response of nanocrystalline metals can be very rate sensitive, so care must be taken not to directly compare property values measured from MD with values from experiments. However, MD simulations with strain rates similar to ours have been shown to capture the important physical processes which govern nanocrystalline plasticity. Schiotz et al. [4] uncovered correlated grain boundary events which produce GB sliding by running MD simulations at a strain rate of $5 \times 10^8$ s$^{-1}$. In addition, Schiotz et al. probed the effect of MD strain rates explicitly and found that pronounced rate sensitivity is only observed for strain rates above $1 \times 10^9$ s$^{-1}$. To explain such observations, Brandl et al. [44] were able to identify a delay in the onset of dislocation propagation as well as reduced levels of cross slip in nanocrystalline Al when strain rates were above $1 \times 10^9$ s$^{-1}$. These authors also observed a noticeable stress overshoot during the early stages of plasticity when such mechanisms were suppressed. Their work revealed that the movement and lifetime of dislocations was highly dependent on the strain rate used to simulate deformation. Since our initial Al sample is dislocation free and we observe very minimal dislocation activity during and after cycling, the propagation or cross slip of pre-existing dislocations is not a key issue. Our simulations mainly report the occurrence of grain boundary accommodated mechanisms (which are expected at such a small average grain size [48]). Moreover, we do not observe a stress overshoot during loading.

A recently developed post-processing algorithm called the Grain Tracking Algorithm (GTA) was used to convert the positions of the atoms in the simulation into structural measurements [33]. This algorithm identifies all crystallites in the system using atomic positions and centrosymmetry parameter (CSP) as required inputs. CSP quantifies the local distortion from a perfectly symmetric crystal lattice at a given atomic position [49] and was used here to distinguish crystalline atoms from those in the GBs. For Al, atoms exhibiting a CSP $\geq 2.83$ Å$^2$ were designated non-crystalline, based on the Gilvarry relation [50] which places an upper limit on the thermal vibrations a crystal can experience without melting. CSP was deemed an appropriate metric for classifying crystallinity during analysis since it is commonly used to identify plastic defects in cubic materials [30]. However, CNA [51] was occasionally used to visualize the grain structure and identify features such as twin boundaries. All visualization of atomic data in this work is performed by the open source particle visualization tool OVITO [31].

Using the face centered cubic (FCC) unit cell as a reference configuration, the nearest neighbor atoms of each crystalline atom are used by the GTA to determine the local crystallographic orientation at each atomic position [33]. Neighboring crystalline atoms are added to grain lists as long as the misorientation measured between their locations is within a designated cut-off angle, $\theta$. If the misorientation angle between crystalline points is above this threshold, the atoms are deemed to belong to a different grain. To capture any low-angle GBs, a strict misorientation cut-off angle of $\theta < 3°$ was used. Once all atoms within a grain are identified, the overall orientation of the grain is found by averaging the calculated local orientations of all atomic points within the grain. This process is continuously repeated until all crystallites within the sample are identified. Figure 1(a) shows the starting microstructure of the 600 K Al sample with all grains identified by color according to the legend below the figure. The black atoms represent boundary or defect atoms which do not belong to any grain. Use of the GTA also allows us to visualize the grain structure in ways that are analogous to the outputs from common experimental characterization techniques. Figure 1(b) projects the {100} poles of each crystallite in the starting configuration onto a plane with a normal along the Z-axis, showing a random texture. Figure 1(c) is a three dimensional crystallographic orientation map of the same sample, with colors corresponding to inverse poles of the grains which are parallel to the Z-axis. Similar to orientation imaging microscopy, which is commonly used to portray data from electron backscattering diffraction (EBSD), the small triangular color-key below Figure 1(c) can be used to identify the orientation of each grain.

With all crystallites and their attributes identified for every simulation time step, evolution of the grain structure was tracked through time. This was accomplished through a mapping of the grain numbers between simulation time steps. In order to map a grain to its past state in a preceding MD output, an atom which lies deep within the interior of the grain is found and its grain number is updated to that of an atom residing near the same coordinates in the preceding time step. All atoms within the grain of interest are then also updated to reflect the proper grain number. This process is repeated for all grains and carried through all MD outputs so that grains are properly tracked for the entirety of deformation.

## 3. Results and Discussion

*3.1 Mechanical Behavior and Cycling Phenomenology*

Overall mechanical behavior was first analyzed by investigating stress-strain curves for the three testing temperatures, as shown in Figure 2(a). Stress was calculated from the system pressure tensor output by LAMMPS, with the Z-component giving the reported uniaxial stresses during cycling. Compared with the monotonic loading behavior, cycled samples demonstrate higher ultimate (peak) strengths, as shown in Figure 2(b). The samples which were mechanically-cycled 10 times were all found to have ultimate strengths at least 10% higher than those of their monotonic counterparts. However, the strengthening from cycling is apparently temporary at the lowest temperature as can be noted by the decrease in flow stress, the stress required to continue plastically deforming the material, at the highest strains. The sample cycled at 300 K exhibits the largest increase in ultimate strength (13.4%), but the flow stress decreases at the highest applied strains to values similar to those of the monotonically strained sample. A combination of elevated temperature and multiple loading cycles seems to ensure that strengthening is more permanent. The black arrows on the right side of Figure 2(a) help illustrate how the flow strength does not return to its initial values at temperatures of 450 K and 600 K.

Further inspection of the mechanical response indicates the presence of hysteresis during cyclic deformation. Because the evolution of hysteresis loops can be a signature of large-scale underlying plastic rearrangement, these loops were analyzed to measure any trends with cycle number. First, the area in each hysteresis loop was measured. Before areas of these loops were calculated, a least squares approximation was applied to the data in order to best fit the small fluctuations in the stress measurements. A Savitzky-Golay filter which fits a series of low-degree polynomials to adjacent data points was used and is a common smoothing technique for decreasing data noise without significant signal distortion. A polynomial order of 3 was used with a frame size of 71, meaning that the data only needed very minimal smoothing [52]. Figure 3(a) is a plot of the 300 K stress-strain response with an overlay of the smoothing function for the first tensile load-unload fatigue cycle. The integrated hysteresis loop area is shown in red for a single cycle to clarify our definition. The hysteresis loop areas are plotted as a function of cycle in Figure 3(b). There seems to be no apparent trend and the areas are relatively constant as a function of cycle.

In a previous high cycle fatigue experiment by Moser et al. [24] regarding the hardening response of nanocrystalline Ni ($d = 23$ nm), the measured mid loop width strains were found to follow two different hardening regimes. The majority of the change they observed in the hysteresis loop size occurred during the first 10 cycles, when GB dislocation sources were active in their material. After exhaustion of these dislocation sources, the mode of deformation shifts to boundary mediated mechanisms and the rate of change of the hysteresis loops decreases significantly. In addition, Moser et al. also discovered that at higher cyclic frequencies the first (dislocation based) plasticity regime can be skipped altogether. Due to the extremely small grain size, lack of dislocations in our samples, and inherently high strain rates of MD simulations it follows that the cyclic plastic deformation of our nanocrystalline Al is taking place in the latter regime, or by collective grain boundary mechanisms, an observation in line with Figure 3(b).

Although the area enclosed by the hysteresis loops does not change, the loops shift down to lower stress values as more cycles are applied. To quantify this change, we measure the change in maximum stress (stress measured at 5% strain) between consecutive cycles and have designated this measure the peak stress difference, $\Delta\sigma$. The reader is again referred to Figure 3(a) in order to clarify our measurement of $\Delta\sigma$. There is an apparent shift in the peak stress difference measured in the first 3 cycles for all temperatures, followed by fluctuation at small values at the higher cycle numbers. The concept that the largest change occurs during the early plastic cycles is consistent with the trends we observe in structural evolution later in this paper. Nevertheless, microstructural rearrangement is taking place as to minimize the overall strain energy, thus reducing the maximum stress attained at 5% as the hysteresis loops shift downward and the metal exhibits hardening. Such cyclic strengthening was also reported by Moser et al. [24] during fatigue experiments of nanocrystalline nickel.

*3.2 Evolution of the Overall Grain Structure*

Having documented obvious changes in mechanical properties, we next investigate the origins of the recorded strengthening through examination of the underlying microstructural evolution. Figures 4(a) through (c) show inverse pole figures with grains labelled according to the number of fatigue cycles. These plots clearly show more grain rotation at the higher temperatures. Some grains were found to merge and/or disappear, leading to an inability to map these grains through the entirety of the simulation. This can be observed in the pole figures as disappearing points (poles that cease to exist).

An increase in grain rotation at elevated temperatures is consistent with the theories for GB plasticity mechanisms that are available in the literature. As theorized by Ashby and Verrall [53], atomic rearrangement through GB sinks and sources allow GBs to act like viscous barriers to the sliding of neighboring crystalline interfaces. A model developed by Harris et al. [54] describes grain rotational rate as a function of net torque and GB diffusivity, implying that the magnitude of grain rotation will have an Arrhenius-type temperature dependence. Moldovan et al. [55] developed an extended model to include arbitrary grain shapes and ended up with an analytic expression involving grain size and temperature dependence that replicates the results of Harris et al. for the specific case of hexagonal grains. Although the driver for net torque acting on a grain can vary between models, all of these theories and others in the literature [17, 56, 57] agree that such rotation is assisted by atomic shuffling within the GBs, implying that increases in temperature should lead to more rotation.

This temperature dependence on grain rotation was analyzed in a quantitative manner using the grain structure information provided by the GTA. Figure 5(a) shows the average rotation for all grains as a function of cycle by comparing the measured orientation after each cycle with the original orientation at cycle 0. It is clear that the higher temperatures allowed for significant increase in average grain rotation. In fact, the average rotation after 10 cycles for the 600 K sample was measured to be 3.7 times that which took place in the 300 K simulation. It is important to note that there were multiple grains which experienced much larger overall rotations than the average. Several grains rotated more than 5° at the highest temperature, especially those grains which shrank throughout the cyclic deformation process.

The GTA data allowed us to measure and track the rotation axis of each grain between cycles as well. It was found that the rotation caused by additional cycles was not occurring about a consistent axis of rotation. As an example, Figure 6 shows the rotation axis as a function of cycle number through the use of an inverse pole figure. This figure tracks the evolution of the rotation axis for a single grain which shrank significantly throughout deformation at 600 K. This sporadic change in axis was found to be characteristic of all grains within the same sample, as well as those in the 450 K and 300 K samples. This observation shows that grains are not rotating in the same direction during each cycle. Rather, they are shifting in different directions through a "ratcheting" type of motion. The monotonically loaded sample was also analyzed in a similar fashion, by tracking rotation axis as a function of strain. Again, the rotation axis changed as a function of strain. Schiotz et al. [4] previously postulated that large rotations about a single axis take place to accommodate and reduce stress, whereas smaller rotations about varying axes help to provide subtle boundary rearrangement. In our analysis, we find no evidence of a common rotation axis for cycled or monotonically loaded samples which suggests that the direction of rotation is determined by reduction in interfacial energy between grains, with applied strain the driver for this mechanism. As a point of caution, we note that the measurement of misorientation axis when rotations are small could be affected by thermal noise as well as an approaching mathematical singularity at the zero point associated with small relative misorientations [58].

The grains with largest measured rotations were the smallest grains in the sample and tended to shrink during the simulation, some of which completely disappeared before the 10$^{th}$ cycle. An example of this is shown in Figure 7(a), which isolates a collection of 5 grains at 600 K and displays their change in orientation from cycle 0. GB atoms are removed from these images which allows for easy three dimensional visualization of grain shape. The red grain in the center rotates significantly during cycling and several boundaries migrate inward as atoms are transferred to neighboring grains. Figure 7(b) shows the relative change in grain size of the 5 grains as compared to their original size measured at cycle 0. There is significant transfer of atoms to the green, purple, and orange grains as the boundaries separating these grains from the central red grain migrate inwards. Cooperative behavior of large rotation with simultaneous GB migration has been simulated by Upmanyu et al. [17] for embedded circular grains with various misorientations. Their results showed that GB migration occurs almost always during rotation except for those boundaries which are highly symmetric. Modeling of the crossover between curvature driven migration and grain rotation induced coalescence was conducted by Moldovan et al. [59], where relative contributions from each mechanism were found to be dependent on the average grain size. Cahn and Taylor [9] also modeled grain shrinking with simultaneous rotation and explained that either coarsening or shrinking could result as long as the total surface free energy is reduced. One of their physical explanations suggest that a migrating boundary allows for the rotation of small groups of atoms from one grain into another. The unit cell becomes distorted by the passing GB and collapses into the neighboring crystal configuration, assuming there is no special symmetry between grains preventing such GB movement. The majority of these models assume simplified microstructures which makes the emergence of these mechanisms even more interesting in our randomly oriented, three dimensional polycrystalline samples.

Deformation-induced grain growth is a common type of structural evolution in nanocrystalline metals. In our simulations, we find coarsening to occur primarily through the process of grain boundary migration and occasionally through coalescence of rotating grains. This resulting coarsening effect is highly temperature dependent and is captured in Figure 5(b), where average grain growth is plotted as a function of cycle. While a very small grain size increase does occur at 300 K ($< 1\%$) and 450 K ($< 3\%$), there is a dramatic increase in average grain size at the highest temperature of 600 K. It is important to note that no significant coarsening occurred during the initial equilibration at 600 K. Hence, the observed grain growth is driven by the mechanical cycling and plastic deformation, albeit aided by the elevated temperature. It is also important to reiterate that while these trends show an increase in average grain size there were in fact many grains which decreased in size throughout the simulation. We again refer to Figure 7 which shows a specific example of the discontinuous grain growth found throughout the Al sample as select grains coarsen or shrink while others keep a relatively constant volume. Discontinuous grain growth of nanocrystalline Al was observed experimentally by Gianola et al. [41] during tensile experiments on nanocrystalline Al thin films ($d$ = 40-90 nm). These authors noted changes in the mechanical properties of the films, namely increased ductility but decreased strength. A previous study by Rupert and Schuh [60] involving repeated frictional wear of Ni-W also provided evidence of discontinuous grain growth, where the grain size distribution broadened with respect to the mean grain size. Similar findings are documented in our 600 K analysis. Of 44 initial grains, 10 showed significant decreases in size during deformation despite the drastic increase in average grain size. This equates to nearly 23% of the grains, several of which disappear completely before the end of the simulation. Because our samples had a very small starting grain size, the increases in strength and required flow stress that we are observing are in part due to inverse Hall-Petch effect [4, 61, 62]. However, the 300 K sample exhibits minimal change in average grain size during cycling and still reaches a higher ultimate stress during the final tensile loading segment. This leads us to believe that there are other microstructural changes taking place which are contributing to the observed changes in mechanical properties.

Grain sliding was also tracked and quantified during cycling. To quantify sliding, we identified the center of mass of each grain and tracked how it moved with respect to each cycle. Figure 5(c) compares the average sliding for all grains within the sample at each of the three testing temperatures. Because the magnitude of this sliding motion is only on the order of angstroms, it is difficult to say that GB sliding lead to significant structural evolution. It is important to reiterate that our definition of grain sliding incorporates the movement of a crystallite's mass center. The trends in Figure 5(c) are similar to those found in Figure 5 (a) and (b), with elevated temperature leading to more evolution. These metrics are likely related, as grain coalescence and coarsening can also shift the mass center. For example, GB migration moves the center of mass of a grain in one direction. Regardless, the magnitude of average sliding is relatively low, even with some contribution from the other mechanisms altering the grain center.

Since grain rotation was significant in the cycled samples, we next investigated to see if there was a trend toward a final texture. Figure 8 presents the final pole figure distribution of grain orientations for {100}, {110} and {111} type poles for all three temperatures at the completion of cycle 10. Comparing the three different testing temperatures, the total number of poles projected decreases at higher temperatures due to the increase in coalescence and disappearance of shrinking grains with increased thermal energy. Grain coalescence allows for a slight shift in the overall projections due to merging of low angle boundaries. However, this coalescence is not drastic enough to extract any conclusive final preferential texture. Any overall changes in the projected poles were seemingly random and sporadic in distribution. This suggests that rotations are occurring in whichever direction is most energetically favorable during each strain cycle, depending on many complex features like compatibility with neighboring grains, and that these rotations lower the energy of the final grain boundary state. Grain rotation as a mechanism of energy reduction is a likely explanation and has been verified in early particle rotation studies [63-65] as well as recent computational work [17], which explains that nano-sized polycrystals will rotate towards local minima in grain boundary energy. Cahn and Taylor [9] have even postulated that coupling between grain rotation and boundary motion can sometimes increase the local interface energy of a grain as long as the configuration results in a decrease in overall system energy.

In an effort to relate these structural changes to the mechanical response of our samples, we compare the trends of Figure 5 with our analysis of the peak stress difference in Figure 3(c). The majority of change in the peak stress difference occurs during the first three fatigue cycles. Additionally, the majority of average rotation and grain sliding has occurred by or before the $4^{th}$ cycle, with the slopes of the trends dampening out even earlier at the two lower temperatures. The remaining grain restructuring that takes place at the higher cycles allows for additional reordering of the GB state through GB migration and changes in the faceting profile between grains. Specific examples of these changes to the grain boundary network will be introduced in the following section.

*3.3 Local Changes to the GB Network*

Local changes to the GB network were next identified and documented. Figure 9 shows cross sections of the final configurations (cycle 10) for each of the three testing temperatures. These cross sections were generated by cutting each sample at the same location and then coloring atoms according to their CNA value. Green atoms are those which have been identified as FCC, white are boundary or defect, and red are those exhibiting the HCP stacking sequence. The final disorientations (minimum misorientations) measured across several grain boundaries are presented in overlaid boxes. The numbers presented are not found by comparing the average orientations of the two grains, but instead by the local disorientation across each boundary. The color of each box corresponds to the absolute magnitude of total disorientation change (denoted $\Delta\Phi$) measured throughout the cycling process. In other words, the total amount of disorientation change measured at the end of cycle 10 with respect to cycle 0.

To begin, it can be visually noted that the GB network has reached a more ordered state at the higher temperatures of 450 K and 600 K. Many GBs have evolved into twin boundaries, or single planes with HCP stacking sequence. To see how the magnitudes of relative grain rotation affect this transition, we look toward the values of $\Delta\Phi$. For the 300 K sample in Figure 9(a), the majority of disorientation changes during cycling were less than one degree, although a few grains exhibit relative rotation between one and three degrees. With such limited reorganization, dramatic structural evolution cannot occur and any changes to the GB structure should be subtle. Previous work has shown that high energy configurations can be relaxed through low temperature annealing or mechanical cycling [66, 67]. Such boundary relaxation can explain the improved mechanical response of the material post cycling and may be occurring in our samples during the first few cycles. Instead of a saturation in dislocation activity taking place like was seen in the previously mentioned fatigue study by Moser et al. [24], it is possible that we are instead observing boundary relaxation of energetic structural regions. Due to our reduced grain size, this boundary relaxation is enhanced through GB plasticity and evidence of this enhanced strengthening response can be noted in Figure 3(c) during the initial cycles.

At the elevated temperatures, the overall relative rotation of grains was found to increase and we see more ordered GBs in Figures 9(b) and (c). In addition, several examples of boundary migration as well as faceted boundaries have helped reorganize the GB network. It is obvious that the most overall change in disorientations occurs in the 600 K sample, where many yellow and red boxes can be found. The red twin boundary at the bottom right of Figure 9(c) labeled "Twin 1" had a disorientation change greater than nine degrees during the simulation. The red box (53° disorientation) above and to the right of this boundary also exhibited a high level of grain rotation in order to accommodate the twinning process that took place in the neighboring grains below. Another twin is identified as "Twin 2." Whether or not these grain pairs reach the twin configuration appears to be highly dependent on temperature and connected to the increasing levels of rotation as temperature is increased. We also make note of the disappearance of a low angle boundary which is identified in Figures 9(a) and (b) but not found in Figure 9(c).

In order to quantify the evolution of the grain boundary state more clearly, we calculated the atomic percent of GB atoms as well as the relative atomic energies of these atoms. Figure 10(a) shows a decrease in grain boundary atomic percent with increasing number of cycles. Twin boundaries were included in this calculation since they separate two different crystal orientations and thus contribute to the grain boundary network. This decrease is partially due to fatigue induced coarsening and coalescence of grains, but there is also an effect of moving from random GBs to twin boundaries in some cases. Twins have very few GB atoms per unit area, because the interface is represented by a single plane of HCP atoms. Figure 10(b) shows the evolution of GB energy with cycling, measured by the excess potential energy as compared to the atoms in the grain interior. Since there is very little dislocation activity present in our samples, their contribution to this calculation is minor. Figure 10(b) shows that the grain boundary network moves to a lower energy state during cycling.

To further illustrate the process of twin formation, Figure 11 highlights eight neighboring grains within the 600 K sample. Several grains of interest have been color coded to help point out the collective microstructural rearrangement that took place during cycling, while the rest of the atoms are colored according to CNA. The random cross section in Figure 11(a) shows the starting configuration of the sample before any cyclic deformation has taken place. Figures 11(b)-(f) show how the microstructure evolves during mechanical cycling, where it is clear that many boundaries have reorganized into low energy configurations. Figure 11(g) presents the calculated misorientation angles between these highlighted grains as a function of cycle number. Multiple GBs converge to a $\Sigma 3$ twin misorientation nearing $60°$ about a <111> type vector, while another (G4-G5) suggests that this trend will continue with an increased number of load cycles. We reiterate that all twins obtained their final configuration through a variety of axes of rotation.

The twin formation between G7 and G8 eventually allows for a reduced misorientation across the original boundary separating G7 and G6. This twinning process, caused by grain rotation as well as the migration of the original separating boundary, ultimately leads to the coalescence of G7 and G6. Further analysis showed that although this grain was indeed identified by the GTA as one grain, the misorientation measured between several points in the grain varies significantly. This is not an error by the algorithm, but instead an example of a bent grain containing low angle grain boundaries. A three dimensional view of this grain post cycling can be seen more clearly in Figure 11(h), where the misorientations between several points along the grain have been identified. CSP identifies all atoms as being crystalline, and the GTA will only find two grains if the local misorientation (misorientation between an atom and its nearest neighbors) is $> 3°$. A similar phenomenon would be difficult to observe in conventional orientation imaging microscopy since these fine grain sizes are near the threshold of such experimental techniques. This is one of a few cases within the sample where two grains merge through a low angle grain boundary. Inspection of this grain shape revealed that the starting Voronoi configuration resulted in a low angle boundary already present within G6 and cycling enlarged this grain through another low angle boundary which initially separated it from G7. Hence, we are left two low angle boundaries within this grain, their approximate locations denoted in Figure 11(h). Analysis of this same grain at the lower temperatures showed considerably less grain rotation and no grain coalescence.

GB faceting as well as the development of coherence across certain twin boundaries was also found to occur more readily with increased temperature. Figure 12 shows an example of twin formation at 600 K, where the grain boundary plane is dynamic. The two grains are segregated by color and all boundary atoms have been removed from the images. In the starting configuration, the boundary between these two grains would already be classified as $\Sigma 3$ according to the well-established Brandon Criteria [68], but the GB is not coherent and the misorientation angle is not exactly 60°. After the first fatigue cycle, the misorientation is increased to almost exactly 60°. Figure 12(b) shows how, at the termination of the first cycle, two different boundary planes exist between these two grains. The interface on the right side is a coherent twin boundary (CTB) with a {111} boundary plane belonging to both crystals. The lower left half is incoherent and thus less energetically favorable. The increased mobility of such an interface facilitates GB migration [69]. When additional fatigue cycles are imposed, the incoherent boundary migrates in a direction parallel to the coherent boundary plane until the entire GB becomes a CTB. This process is highlighted in Figures 12(b) and (c) with black arrows signifying the direction of migration. The remaining cycles allowed for rearrangement of the faceting profile between the twinned grain pair as can be seen in Figures 12(d) and (e). The number of facet steps decreased with additional cycles, suggesting that the faceted structure is only an intermediate GB state that facilitates the transition to a perfect twin between the grains.

Based on these observations, we attribute the permanent increase in mechanical flow strength after high temperature cycling (as shown in Figure 2(a)) to the higher levels of structural evolution. The formation of low energy twin boundaries as a result of grain rotation is likely a major contributor to this effect. It is not uncommon in larger grained nanocrystalline metals for twins to have a significant impact on strength and ductility with many researchers striving to grain boundary engineer (GBE) metals by increasing twin densities [70, 71]. However, in nanocrystalline metals with fine grain sizes close to what was explored here, traditional deformation twins are rare and an inverse grain-size effect on twinning accounts for the higher levels of generalized planar fault energy needed to emit twinning partials [72]. The formation of special boundaries observed here could have important implications regarding GBE of nanocrystalline materials. Direct twinning by rotation, GB migration, and sliding may provide a path for increasing the fraction of special boundaries in a nanocrystalline grain structure, without the traditional GBE processes of dislocation network formation and recrystallization.

## 4. Conclusions

Quantitative characterization of nanocrystalline structural evolution due to novel GB deformation mechanisms is important for designing against fatigue and for grain boundary engineering. Because these plastic processes can be extremely difficult to observe even with costly in-situ experiments, we have used molecular dynamics simulations to observe and document trends in structural evolution during low cycle fatigue. The results shown here provide a quantitative analysis of several deformation mechanisms pertaining to nanocrystalline metals with extremely small average grain size. The following conclusions can be drawn from this study:

- Cyclic strengthening is observed and the magnitude of ultimate strength increase is dependent on the number of cycles. We attribute this strength change to the structural evolution which is driven by plasticity. Increased testing temperature ensures that this strength increase is permanent and persists to larger applied strains, by causing more obvious changes to the GB network.
- Both grain rotation and grain boundary migration were observed for all samples and can contribute to discontinuous grain growth as atoms are transferred between grains. Another byproduct of such growth is the shrinking and disappearance of many grains, with these grains often exhibiting the largest overall rotations.

- Grain growth was also found to occur through the coalescence of neighboring grains as misorientation between grains was reduced with more cycles and elevated testing temperature. This mechanism for energy reduction allowed for the formation of elongated grains consisting of several low angle GBs leading to the formation of bent grains.

- Grain sliding was tracked quantitatively and found to be temperature dependent, but ridged body translation of grains was not identified as being a dominant deformation mechanism. The overall magnitudes of grain sliding were found to be relatively small, on the order of a few angstroms.

- A significant number of twin boundaries were formed in the specimen cycled at 600 K as a result of collective deformation physics. Elevated temperature allows more rotation to occur by aiding GB diffusion, with rotation continuing until misorientation approaches the 60° misorientation angle of a $\Sigma 3$ twin.

- The majority of evolution caused by grain rotation occurs in the first few cycles with additional cycles leading to a more ordered final GB state with fewer defect atoms. Boundaries continue to migrate and begin to reorganize as twin boundary planes become more coherent with cycling.

This study has provided insight into the collective nature of GB-dominated deformation mechanisms which take place in nanocrystalline metals, with a specific focus on adding quantitative understanding. The ability to quantify relative contributions of each mechanism is necessary in order to identify any emerging trends in structural evolution during nanocrystalline plasticity. Since all grain structure attributes are available in an atomistic simulation, interesting and important individual examples of plasticity can be easily identified within the sample and explored in depth. These capabilities are what make this type of analysis a powerful tool for understanding the underlying physical processes that take place in nanocrystalline metals.

## Acknowledgments

We gratefully acknowledge support from the National Science Foundation through a CAREER Award No. DMR-1255305.

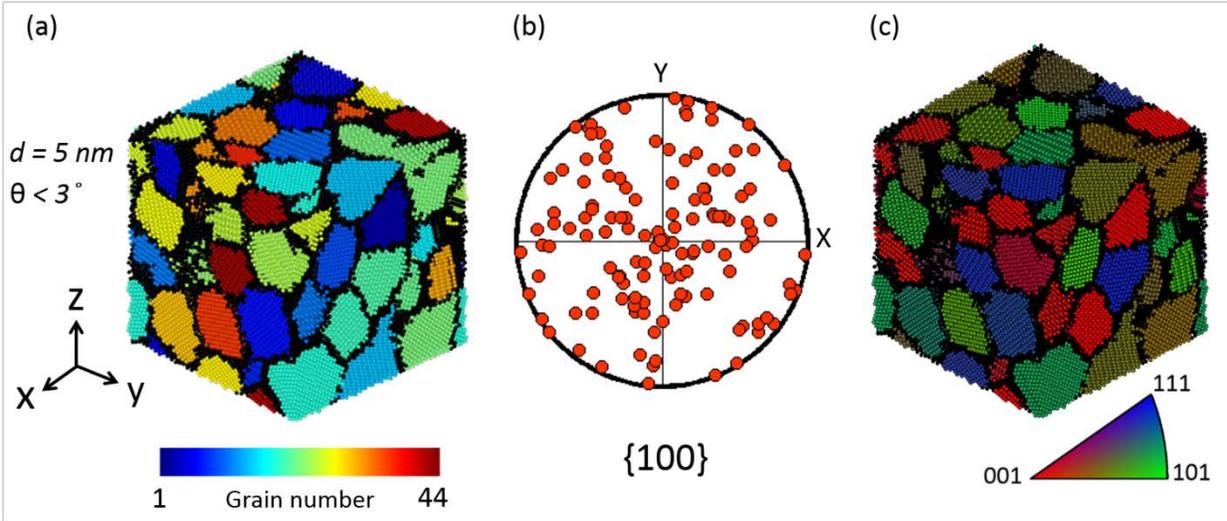

*Figure 1.* (a) Starting configuration of the polycrystalline Al sample with 44 grains, each identified by color. Automation of grain identification was done using the Grain Tracking Algorithm with a specified misorientation cutoff of $\theta < 3°$ between crystalline nearest neighbors. Black atoms are those which were identified as GB or "other" atoms. (b) Pole figure representation of overall random texture distribution of all grains. (c) {001} Orientation map showing distribution of all grain normals with respect to the simulation Z-direction. The reader is referred to the online version of this article for color figures.

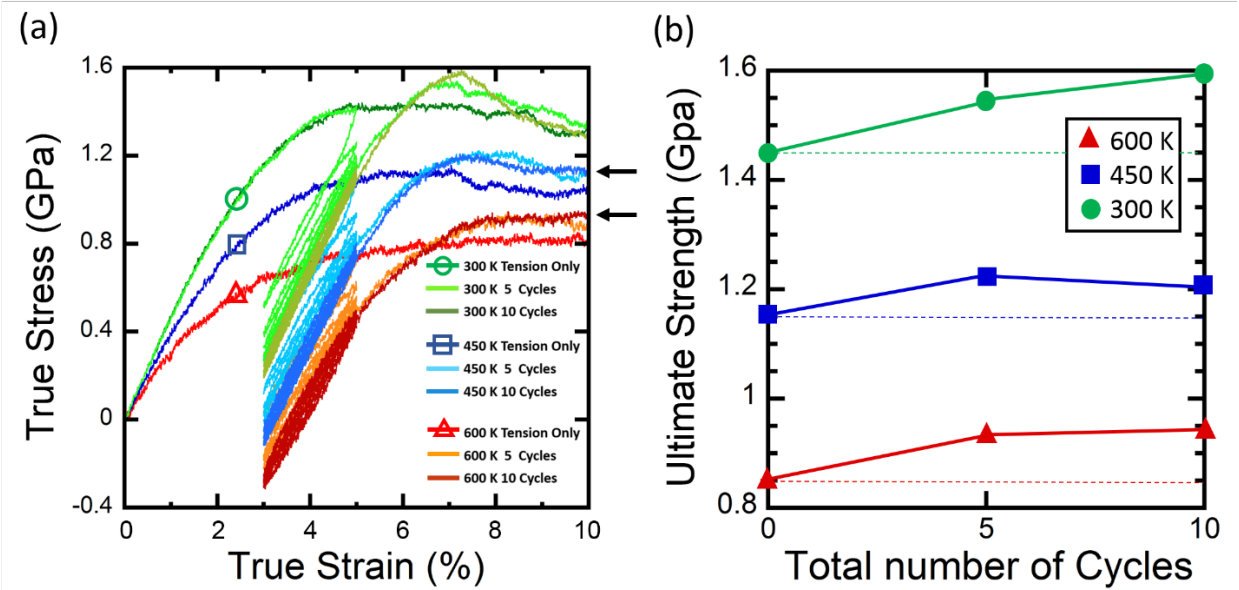

*Figure 2.* (a) Results of monotonic tension and tensile load-unload fatigue simulations. All tests were strain controlled with the same initial starting configuration. (b) Ultimate (peak) strength as a function of cycle number for all three fatigue conditions (0, 5, or 10 cycles). Dotted lines are drawn to help illustrate the level of strengthening observed.

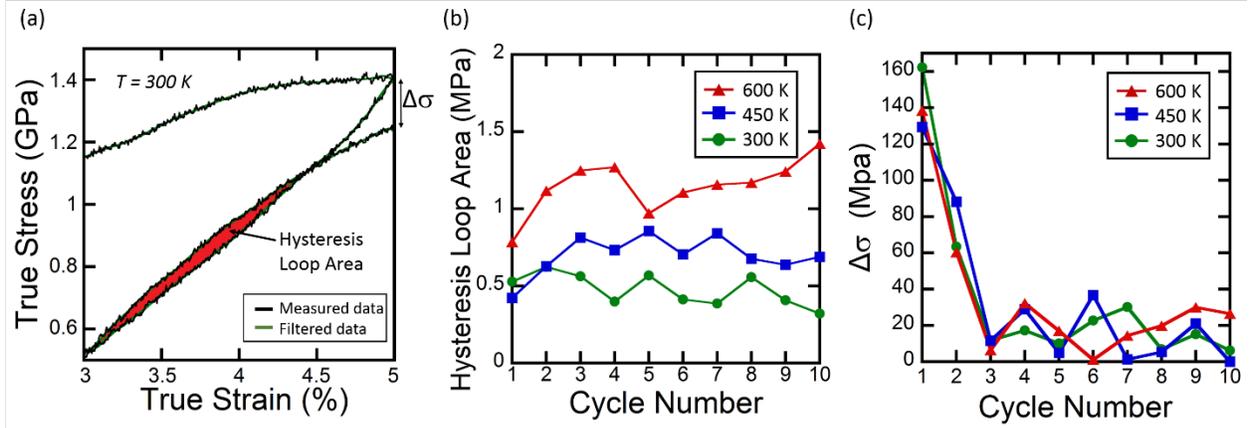

*Figure 3.* (a) A smoothed portion of the 300 K stress strain response illustrating a hysteresis loop area as well as the peak stress difference (Δσ) measurement for one complete loading cycle. (b) Enclosed hysteresis loop areas for all temperatures as a function of cycle number. (c) Peak stress difference (Δσ) measurements as a function of cycle number.

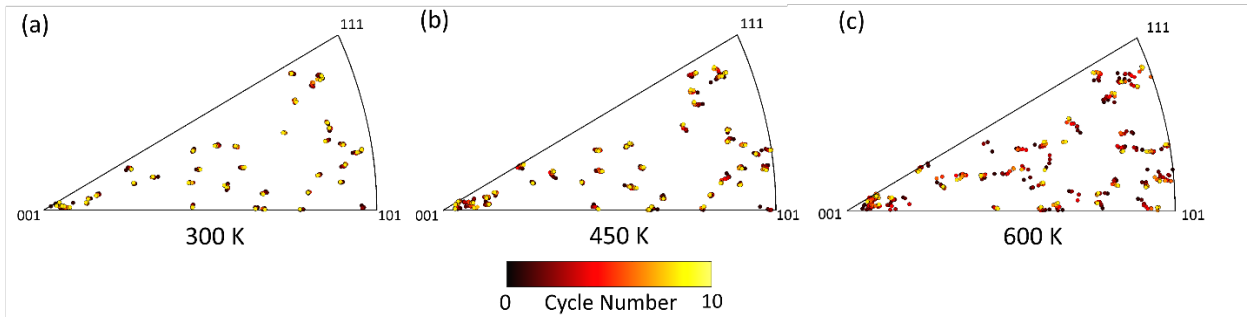

*Figure 4.* Z-Direction inverse pole stereographic projections show the evolution of the orientations of all grains at each temperature. Displacement of each point gives a visual representation of grain rotation caused by each cycle.

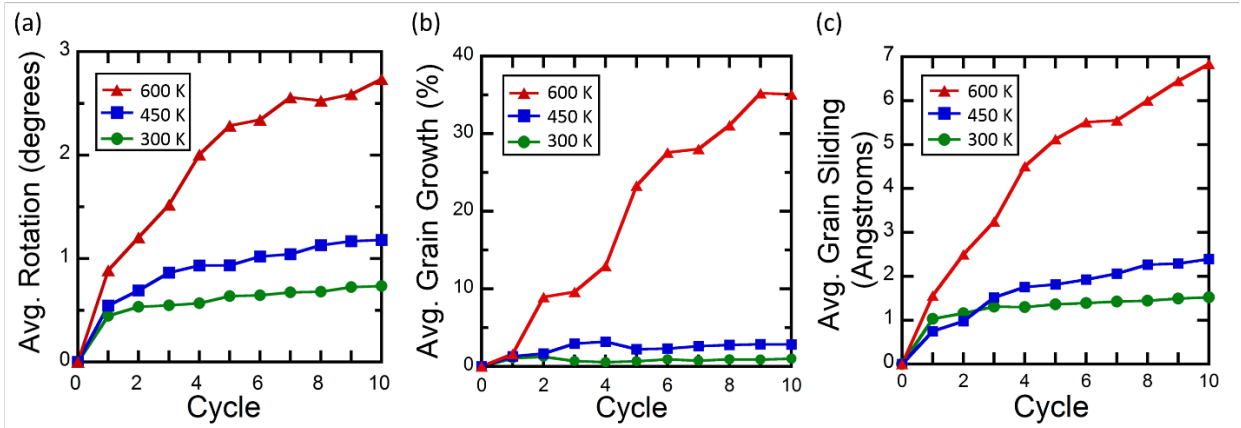

*Figure 5.* (a) Average rotation of all grains with respect to their original configuration at cycle 0. Rotation was measured by calculating the misorientation angle between the original orientation and the new orientation found after each load cycle. (b) Grain growth as a function of cycle for all three temperatures. Average grain size calculated after each cycle was compared to that which was found at cycle 0. (c) Average grain sliding as a function of cycle number. The displacement of the center of mass of each grain was found and compared with the original location at cycle 0.

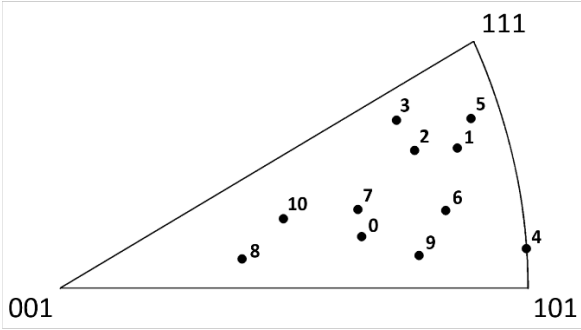

*Figure 6.* Stereographic triangle showing the random evolution of the misorientation axis for a single rotating grain as a function of cycle number.

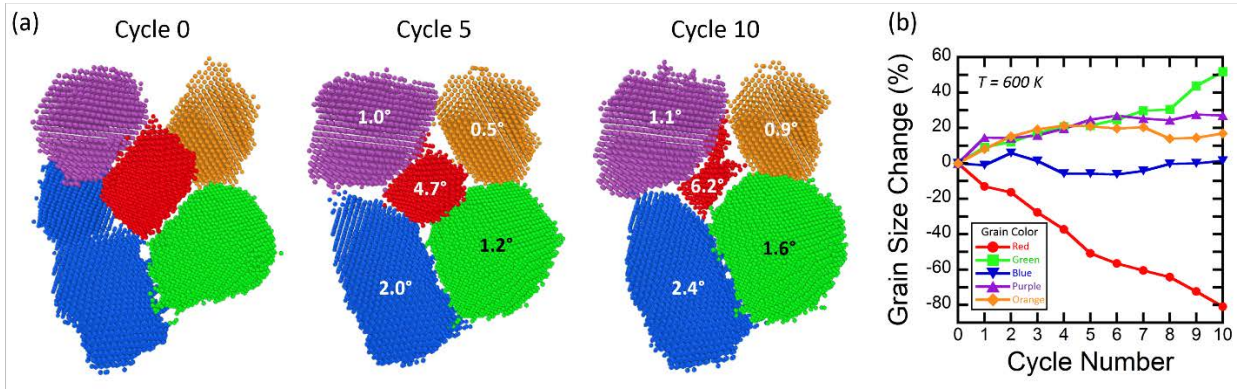

*Figure 7.* (a) Tracking of 5 different grains during 600 K cycling. Images show final arrangement of grains at the end of each cycle along with magnitude of rotation from original configuration at cycle 0. (b) Grain size evolution measured as percent difference from size measured at cycle 0 for the same 5 grains.

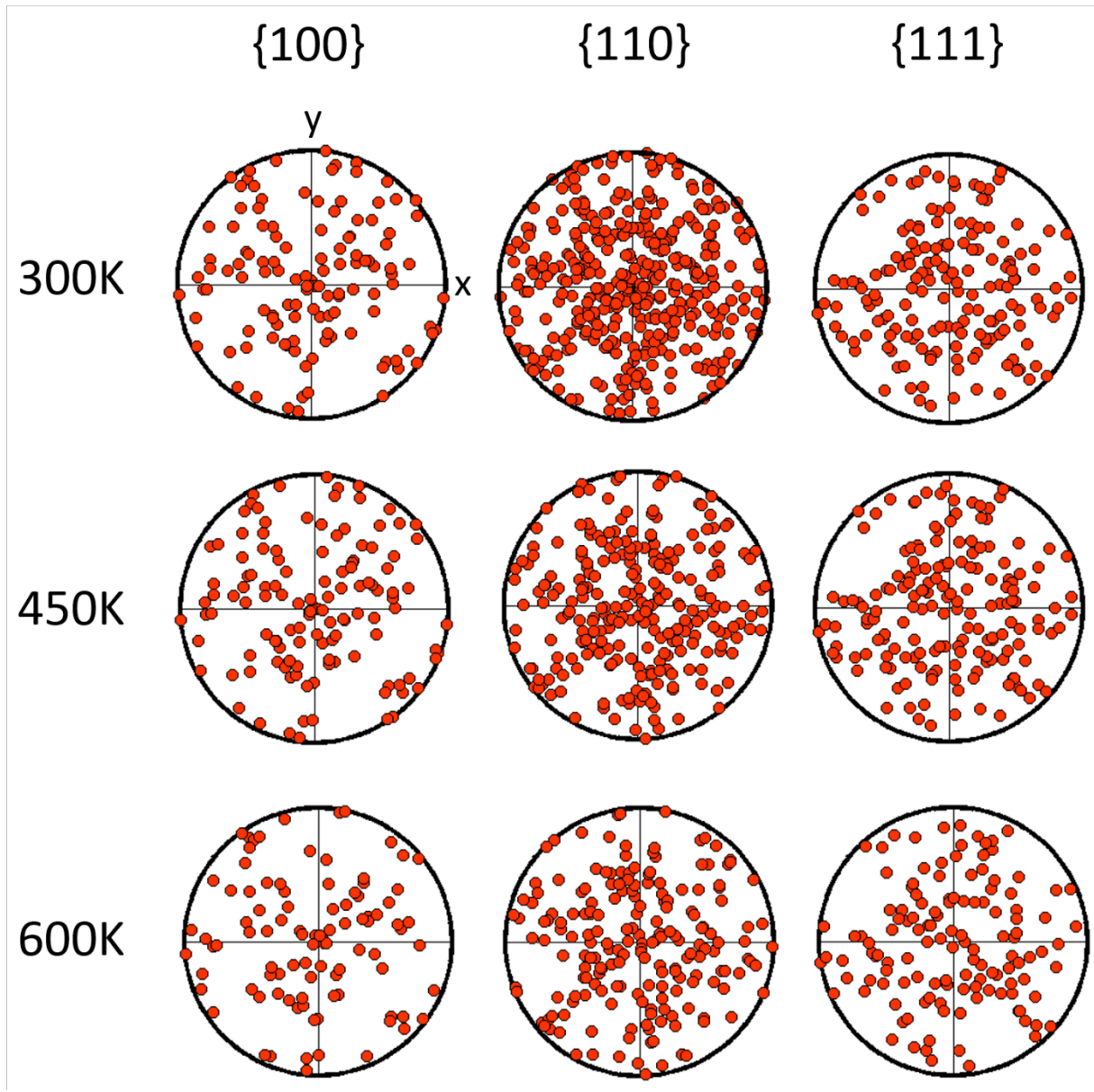

*Figure 8.* Final pole figures for all three samples taken after the completion of cycle 10. There is no discernable evolution toward a global texture.

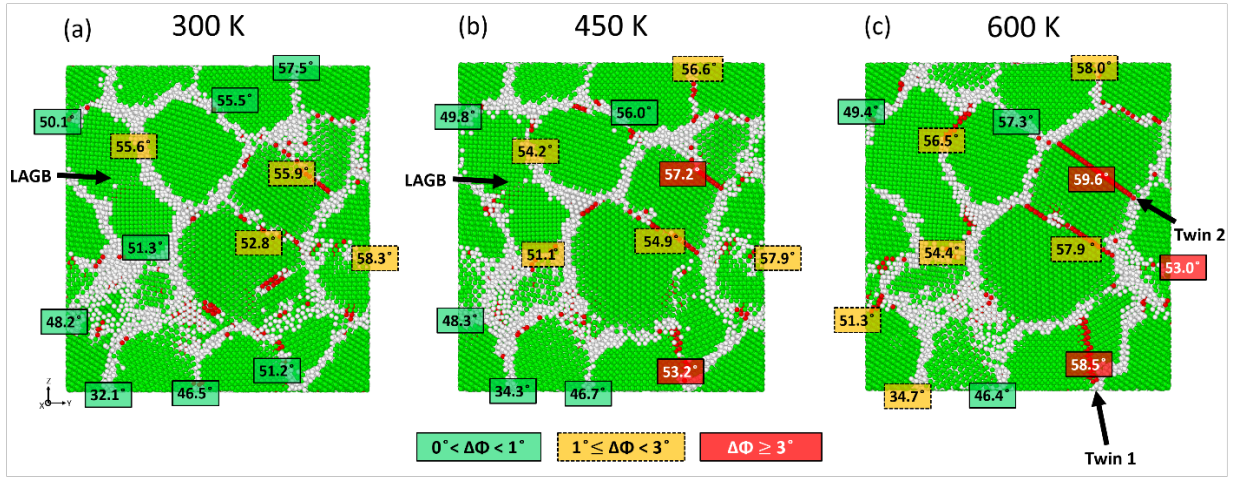

*Figure 9.* Cross sections showing the final microstructures of the (a) 300 K, (b) 450 K, and (c) 600 K Al samples after 10 fatigue cycles. Final disorientations measured across several boundaries are given in the small boxes with the color of the box representing the magnitude of the overall disorientation change (ΔΦ) accumulated since cycle 0.

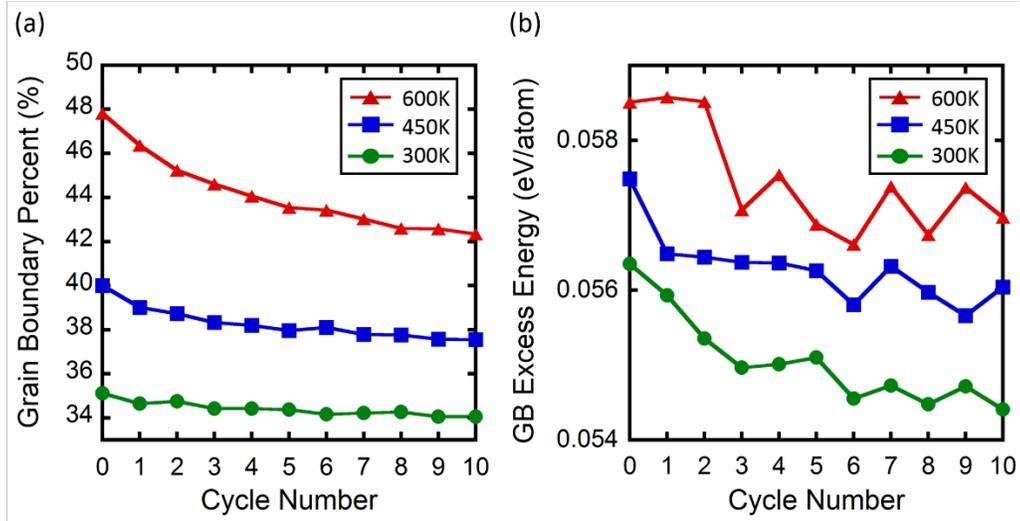

*Figure 10.* (a) Percentage of grain boundary atoms as a function of load-unload cycle. (b) Grain boundary excess energy as a function of load-unload cycle.

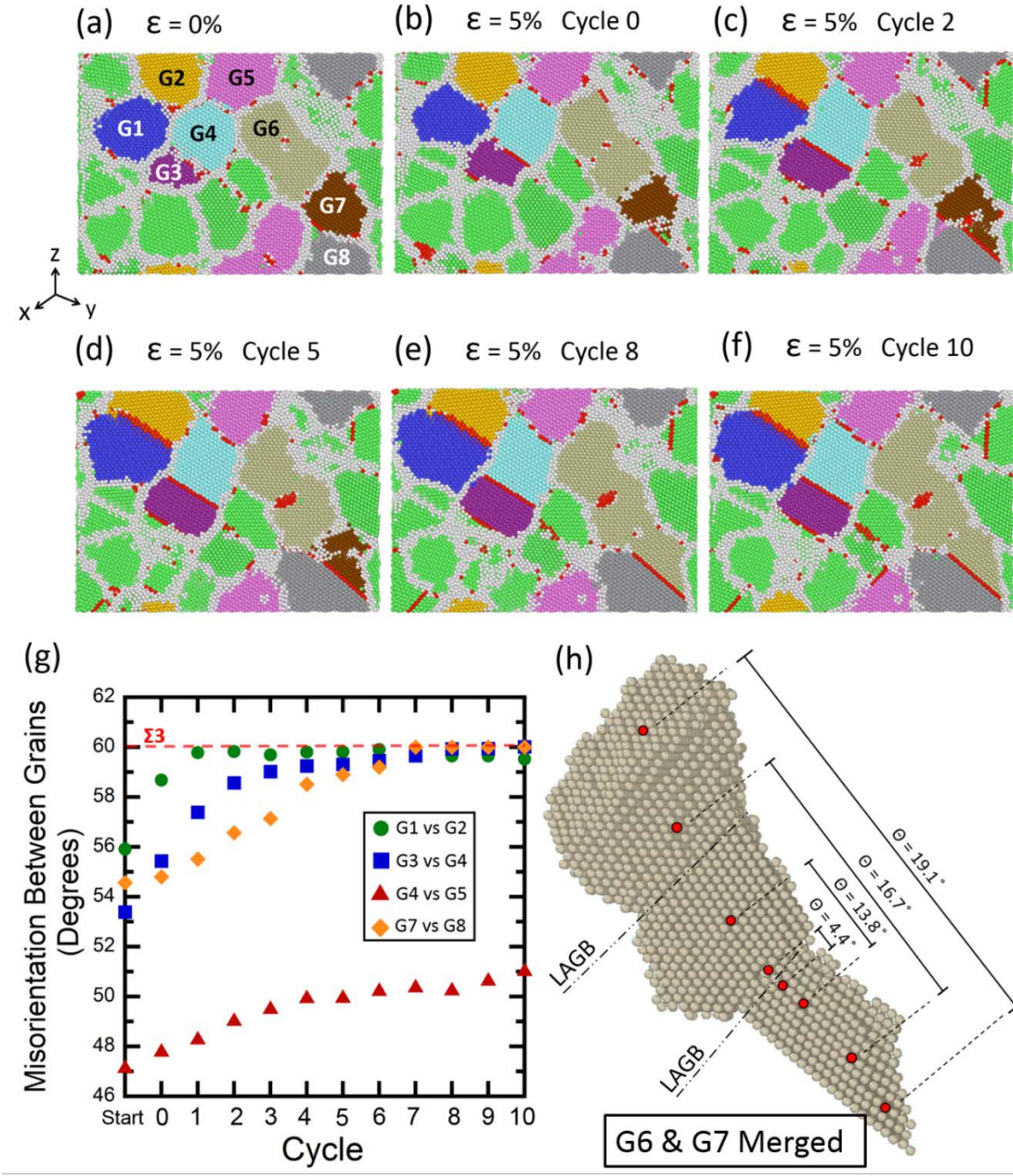

*Figure 11.* (a) Snapshot of a cross section of the starting configuration for the 600 K cycled Al sample with grains numbered and colored for clarity. (b)-(f) The final configuration after select cycles, showing the evolution of microstructure of several grains. Rotation of grains allows for formation of several deformation twins. (g) Plots the misorientation between certain grains and shows that many of the grain pairs are converging to a 60° misorientation of a Σ3 twin boundary. (h) A three dimensional view of G6 and G7 after merging during cyclic loading. Misorientation between several points along this grain illustrates the bent configuration of this new crystallite.

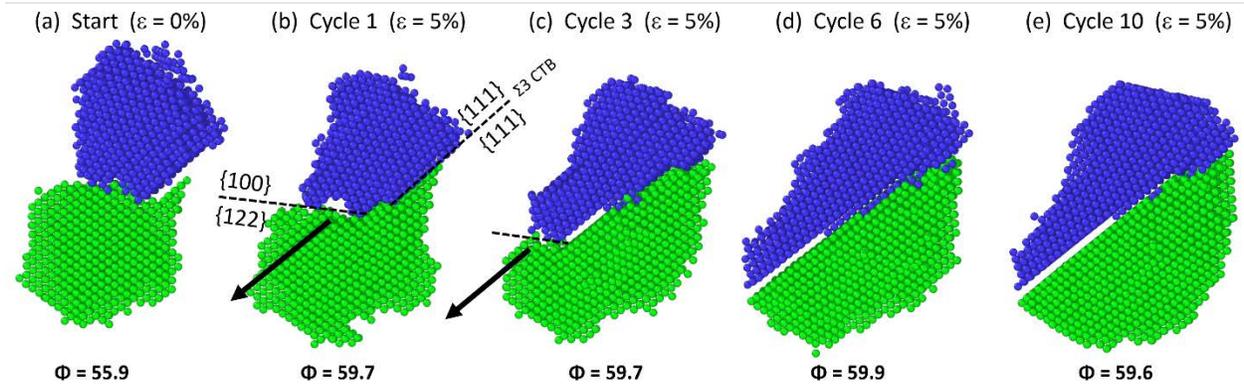

*Figure 12.* (a) Starting configuration of a select grain pair at 600 K with average misorientation listed below the image. (b) GB character of the twin boundary interfaces at the completion of Cycle 1 and start of incoherent boundary migration. (c) Continued migration of the incoherent twin boundary across the grain. (d) – (e) subtle changes in interface structure and GB faceting profile.